# Comparison of Sn-doped and nonstoichiometric vertical-Bridgman-grown crystals of the topological insulator $Bi_2Te_2Se$

S. K. Kushwaha[1], Q. D. Gibson[1], J. Xiong[2], I. Pletikosic[2,3], A. P. Weber[4], A. V. Fedorov[5], N. P. Ong[2], T. Valla[3] and R.J. Cava[1]

[1]*Department of Chemistry, Princeton University, Princeton NJ 08544, USA*

[2]*Department of Physics, Princeton University, Princeton NJ 08544, USA*

[3]*Condensed Matter Physics and Materials Science Department, Brookhaven National Lab, Upton, New York 11973, USA*

[4]*National Synchrotron Light Source, Brookhaven National Lab, Upton, New York 11973, USA*

[5]*Advanced Light Source, Lawrence Berkeley National Laboratory, Berkeley, California 94720, USA*

A comparative study of the properties of topological insulator $Bi_2Te_2Se$ (BTS) crystals grown by the vertical Bridgeman method is described. Two defect mechanisms that create acceptor impurities to compensate for the native *n*-type carriers are compared: Bi excess, and light Sn doping. Both methods yield low carrier concentrations and an *n-p* crossover over the length of the grown crystal boules, but lower carrier concentrations and higher resistivities are obtained for the Sn-doped crystals, which reach carrier concentrations as low as 8 x $10^{14}$ $cm^{-3}$. Further, the temperature dependent resistivities for the Sn-doped crystals display strongly activated behavior at high temperatures, with a characteristic energy of half the bulk band gap. The (001) cleaved Sn-doped BTS crystals display high quality Shubnikov de Haas (SdH) quantum oscillations due to the topological surface state electrons. Angle resolved photoelectron spectroscopy (ARPES) characterization shows that the Fermi energy ($E_F$) for the Sn-doped crystals falls cleanly in the surface states with no interference from the bulk bands, that the Dirac point for the surface states lies approximately 60 meV below the top of the bulk valence band maximum, and allows for a determination of the bulk and surface state carrier concentrations as a function of Energy near $E_F$. Electronic structure calculations that compare Bi excess and Sn dopants in BTS demonstrate that Sn acts as a special impurity, with a localized impurity band that acts as a charge buffer occurring inside the bulk band gap. We propose that the special resonant level character of Sn in BTS gives rise to the exceptionally low carrier concentrations and activated resistivities observed.

## I. INTRODUCTION

Topological insulators (TIs) are a new phase of quantum matter where strong spin-orbit coupling and band inversion, critical aspects of the bulk electronic structure of certain small band gap semiconductors[1,2], result in the presence of metallic electronic states with a Dirac-like energy dispersion on the crystal surfaces. These "topological surface states" are found on a handful of materials.[3-5] The prototype tetradymite-type materials $Bi_2Se_3$ and $Bi_2Te_3$ are suitable for many studies, but their bulk conductivities are too high to allow for the study of the transport properties of the electrons in the surface states without interference from the bulk electrons. The ternary variant $Bi_2Te_2Se$ (BTS) has emerged as a bulk material of choice for study of the charge transport of electrons in the topological surface states without severe interference from bulk conduction. This distinct compound has a well ordered structure, consisting of stacked [Te-Bi-Se-Bi-Te] quintuple layers, and, although it has a small band gap (~300 meV), it has been grown as bulk crystals with relatively low carrier concentration. As-grown BTS crystals exhibit metallic conduction with a high *n*-type career density of $\sim 10^{19}$. The tailoring of atomic composition results in more resistive BTS, and bulk crystals with varied atomic compositions have been reported.[6-9] Our previous studies have shown that a small excess of Bi in BTS results in $Bi_{Te}$ antisite defects, which act as acceptors and compensate the native source of electrons, resulting in crystals with carrier densities of $\sim 10^{16}$ $cm^{-3}$ and relatively high bulk resistivities.[7] High resistivities are also achieved for the complex quaternary $Bi_{2-x}Sb_xTe_{3-y}Se_y$ solid solutions; these crystals display significant lattice disorder,[6] which influences topological surface state transport.[10-12] Alternatively, Sn acts as an acceptor in BTS and donates holes that can compensate the native donors; enhancement of the bulk resistivity at low temperatures and low carrier concentrations for Sn-doped BTS crystals grown by slow furnace cooling has been observed.[13,14]

It is well known that improving the quality of crystals is crucial for realizing the optimal electronic properties in materials[15] and therefore it is of importance to determine methods for growing higher quality bulk single crystals of existing TIs. The promise that BTS has already shown as a host for experiments on topological surface states suggests that it is an excellent candidate material for optimization. Of primary interest is to focus on the growth of high resistivity bulk crystals of BTS with the lowest possible bulk carrier concentrations, such that the Fermi level lies in the bulk band gap, the energy regime where the surface states lie. Except for a few reports, however, little specific attention has been paid to the growth of high quality bulk single crystals of TI materials.[7-9,16] In addition, it has been reported that during some bulk BTS crystallization processes, phase separation is a serious problem, which can result in the formation of metallic and insulating regions in proximity in the same crystal boule.[7,16,17] It is therefore of interest to develop a crystal growth process that results in homogeneous, insulating, low carrier concentration bulk BTS crystals. Furthermore, it has been mostly assumed until now that dopants merely change the carrier concentration in TI crystals through normal defect-chemistry processes; effects of special dopants on the electronic structure have not been considered.[18]

In view of the above, here we report the growth of two types of carrier compensated BTS crystals by the vertical Bridgman technique (VBT), employing a specially designed furnace. The compensation was achieved by inducing $Bi_{Te}$ antisite defects, i.e. in $Bi_{2.04}Te_{1.96}Se$, and alternatively by 0.5% Sn doping in stoichiometric BTS, i.e. in $Bi_{1.99}Sn_{0.01}Te_2Se$. The long boules from the Bridgeman growth were sectioned and their resistivities and bulk carrier concentrations were studied. Both types of crystals are highly resistive at low temperatures, with carrier densities as low as $8\ x10^{15}$ and $8\ x\ 10^{14}\ cm^{-3}$ for the Bi-excess and Sn-doped crystals, respectively. The Sn-doped crystals display resistivities greater than 10 Ωcm and thermally activated resistivities for large regions of the crystal boule that are characteristic of half the bulk band gap. Angle resolved photoelectron spectroscopy (ARPES) data confirm the presence of the high quality surface states, the bulk band gap of 300 meV, and the position of the surface Dirac point about 60 meV below the top of the bulk valence band. The ARPES data further allow for the determination of the carrier concentration as a function of energy for both the surface and bulk electronic states with respect to the top of the valence band. Both the Bi excess crystals and the Sn doped crystals exhibit Shubnikov-deHass (SdH) oscillations from the topological surface states, but those in the Sn-doped crystals are particularly well defined, revealing that the Sn-doping does not interfere with the presence of high quality surfaces on BTS crystals. Electronic structure calculations demonstrate that an Sn impurity in BTS behaves in a way that is distinct from a $Bi_{Te}$ antisite defect, leading us to conclude that localized Sn states act as a carrier concentration buffer, pinning the Fermi energy in the gap, explaining why Sn doping allows for crystals with much lower carrier concentrations and higher resistivities than those with Bi excess.

## II. EXPERIMENTAL

The bulk single crystals of non-stoichiometric $Bi_{2.04}Te_{1.96}Se$ (BTS), and Sn-doped $Bi_{1.99}Sn_{0.01}Te_2Se$ (BTS:Sn) were grown by the vertical Bridgman technique (VBT). These compositions were selected for detailed study due to their expected proximity to the *n*-*p* crossover in the BTS system.[9] A tubular furnace of 2.5 cm inner diameter and 100 cm length with good thermal insulation was employed. Two thermal zones were installed to obtain the required temperature profile. Along the furnace height, the temperature in the upper region (~40 cm) was maintained well above melting point of BTS, and in the crystallization region a high thermal gradient (20 K/cm) was created between the hotter and cooler zones.

Before the crystal growth, the high purity (5N) metals Bi and Te were reduced by heating under vacuum (~$10^{-3}$ mbar) in the presence of C for 12 hrs at 1173 K, and 5N pure Se was further purified by recrystallization. The required amounts of purified elements (~ 27 gm in total) were filled into carbon-coated pointed bottom quartz ampoules (6 mm in inner diameter) under an Ar atmosphere in a glove-box. For the Sn-doped crystal, the required amount of 5N pure Sn was added. The ampoules were sealed (tube length 27 cm) under a high vacuum of ~$10^{-5}$ mbar; before sealing, the ampoules were flushed with Ar multiple times to avoid the presence of traces of air in the growth tube. The sealed ampoules were heated at 1173 K for 12 hours, and stirred to insure the homogeneous melting of the charge and avoid bubble formation at the bottom of the ampoule. After melting and homogenizing the charge at 1173 K, the temperature of the furnace was slowly lowered such that the temperature of the lower, low temperature zone was ~973 K while the upper high temperature zone was at ~1023 K. Both of these temperatures were held constant for the growth process. The ampoules were slowly lowered through the crystallization zone, at the translation rate of 1 mm/hr, while rotating at 2 rpm. The pointed bottom of the ampoules helps to avoid multi-nucleation, and hence facilitates the growth of a single crystal over a long length of the ampoule. On completion of the growth process, the temperature of both zones was slowly lowered (10 K/hr) to 773 K to anneal the crystal for 24 hours. This was followed by cooling to room temperature at the rate of 120 K/hr. The growth parameters were identical for both crystals. High quality crystal boules of ~ 13 cm length were successfully harvested for both growths. The *c*-axis of both types of crystals was oriented perpendicular to the crystal growth direction. The boules were cut into pieces of 1-2 cm length for characterization purposes. The specimens thus obtained of BTS and BTS:Sn were designated by the letters A to H and A1 to F1, respectively, with the alphabetical sequence running from the first to freeze to the last to freeze sections.

The crystal structure of the solidified materials was confirmed by powder X-ray diffraction for pieces from the first, middle and last to freeze sections of the boules. The BTS crystal structure was observed, with no change in lattice parameter and no additional phases. The crystals were easily cleaved along the hexagonal (001) crystallographic plane to make specimens for investigating the temperature dependent bulk resistivities ($\rho$) and carrier densities. The transport measurements were performed on a Quantum Design Physical Property Measurement System (PPMS) in the temperature range of 10 – 300 K. The resistivities were measured by the linear four point probe method and the carrier densities were evaluated by the five point probe Hall-effect measurement method. For the Hall measurements, a constant magnetic field of 1 Tesla was applied normal to the sample (001) plane. The electrical contacts on the freshly cleaved surfaces were made by silver paint through annealed platinum wire. The magnetoresistance data were measured in an American Magnetics 15T magnet, similar to previous studies.[12]

The ARPES experiments were carried out on a Scienta SES-100 electron spectrometer at beamline 12.0.1 of the Advanced Light Source. The spectra were recorded at photon energies ranging from 30 to 100 eV, with a combined instrumental energy resolution of ~12 meV and an angular resolution better than ±0.07°. Samples were cleaved at low temperature (15- 20 K) under ultra-high vacuum conditions (base pressure better than $2\times10^{-9}$ Pa). The temperature was measured using a silicon sensor mounted near the sample.

Electronic structure calculations were performed using density functional theory through the Wien2k code[19] with a full-potential linearized augmented plane-wave and local orbitals basis together with the Perdew-Burke-Ernzerhof parameterization of the generalized gradient approximation.[20] The wave cutoff parameter RMTKmax was set to 7 and the Brillouin zone (BZ) was sampled by 100 k-points. Spin-orbit coupling (SOC) was included. For all calculations, a 2x2x2 supercell of the hexagonal $Bi_2Te_2Se$ cell was used, with a total of 120 atoms per cell, with overall symmetry P-3m1. For the Sn doped case, 2 out of the 48 Bi atoms were replaced by Sn, which simulates an approximately 4% doping level. Similarly, for the $Bi_{Te}$ antisite defect calculation, 2 out of 48 Te atoms were replaced by a Bi.

## III. RESULTS AND DISCUSSION

### A. Resistivity and carrier density

Figure 1(a) shows the $\rho$ vs. $T$ behavior in the temperature range of 13 K to 290 K for the eight crystal specimens cut from different positions of the Bi-excess BTS crystal boule. At higher temperature, up to 150 K, $\rho$ is small for all the specimens. The room temperature $\rho$ value varies in the range from 5 to 35 mΩcm. As the samples are cooled below 150 K, all show a significant increase of $\rho$, and a long length of the boule shows low temperature resistivities in the 1 - 2 Ω-cm range, the state of the art for BTS.[6-9] The increase in $\rho$ on cooling for all the samples indicates nonmetallic behavior, but the increase in resistivity on cooling for the first to freeze samples cut from the boule (A and B) is very small. The resistivity for cooled crystals from the middle sections C, D and E starts increasing at relatively higher temperatures and becomes saturated at low temperature. Saturation in resistivity at low temperatures is unusual for traditional semiconductors, and has been attributed to the presence of a metallic "short circuit" component contributed by the conduction of the surface states in topological insulators.[21] The samples (F, G and H) from the last to freeze region of the boule do not show such saturation of the resistivity. The inset of Figure 1(a) shows log $\rho$ vs. 1000/$T$ Arrhenius plots, for T > 100 K, characterizing the thermally activated resistivity region. The activation energy ($\Delta$), is evaluated by using the activation law $\rho \sim \exp(\Delta/K_B T)$, where $K_B$ is the Boltzmann constant. The values of $\Delta$ are found to be 24, 27 and 20 meV for samples C, D and E, respectively, which are slightly higher than those reported for Se-rich furnace-cooled solid solution BTS single crystals.[6] These values likely reflect the presence of an acceptor level about 25 meV above the valence band in the nonstoichiometric BTS crystals.

The corresponding $\rho$ vs. $T$ plots for the Sn-doped boule, BTS:Sn, are shown in Figure 1(b). It can be seen immediately that these are qualitatively different from those of the Bi-excess BTS crystal. The low temperature $\rho$ values commonly fall in the 2-10 Ω-cm range, and reach that value at much higher temperatures than is seen in the Bi-excess BTS. This high resistivity is obtained for a long section of the boule, ~6-8 cm in length. At high temperatures $\rho$ is also high, and for most of the samples ranges from 50 - 500 mΩ-cm. The samples ranging from first to freeze to the middle of the boule (A1, B1, C1, D1 and E1) show saturation in $\rho$ over a large range of temperature, implying the likely presence of a larger relative surface electron contribution to the conduction. The log $\rho$ vs. $1/T$ plots are shown in the inset to Figure 1(b) to show the thermal activation. The Arrhenius law fits well to the high temperature data, and the evaluated $\Delta$ values for samples A1, B1, C1, D1, and E1 are found to be 116, 135, 144, 113, and 109 meV, respectively. The $\Delta$ values are quite high; B1 and C1 exhibit the highest $\Delta$ values (135 and 144 meV) thus far reported for BTS though they are in the range of those reported for Sn-doped BTS crystals extracted from slow cooled melts.[13] These values are very close to half to the bulk band gap $E$g (~ 300 meV) of an ideally perfect BTS and indicate that the bulk Fermi level lies in the bulk band gap.[18] The shallow impurity levels that normally dominate the resistivity are not observed. The high $\rho$ and $\Delta$ values suggest that vertical-Bridgeman-grown BTS:Sn is an excellent candidate for studying surface state transport phenomena in topological insulators.[22]

The temperature dependent Hall resistivity ($R_H$) measurements for the Bi-excess BTS crystals are shown in Figure 2(a). At high temperature, the $R_H$ values are small for all the samples, but as the samples cool below 100 K, $R_H$ increases to high values. Samples A and B have very low negative values over the entire temperature range, characteristic of a high density of *n*-type carriers. Samples C, D and E exhibit positive $R_H$ values at high temperature, and, as the temperature lowers below 50 K, $R_H$ peaks at about ~40 K and crosses to negative values at ~ 30 K, indicating a change in the dominant carrier type from *p*-type to *n*-type. The negative $R_H$ values increase sharply with decreasing temperature and reach ~1300 $cm^2/C$, which implies a low carrier density (~ 8 x $10^{15}$ $cm^{-3}$). The samples F, G and H, from the last-to-freeze part of the boule, have positive $R_H$ over the entire temperature range. Figure 2(b) shows, similarly, the $R_H$ vs. T plots for the BTS:Sn crystal. These plots look quite different from those of the Bi-excess BTS. The samples A1-D1 show negative values of $R_H$, however, sample B1 shows the change of sign of $R_H$, exhibiting the carrier crossover with temperature. Sample C1, near the middle of the boule, exhibits different behavior from the others as it has a high negative $R_H$ over the entire temperature range, and displays its highest $R_H$ value (7050 $cm^2/C$) at ~150 K. This $R_H$ value corresponds to a carrier density of $\sim 8\times10^{14}$ $cm^{-3}$. As for the non-stoichiometric BTS boule, the last to freeze sections of the BTS:Sn boule are *p*-type for the full temperature range, but a large portion of the crystal is *n*-type.

For BTS crystals near the *n-p* crossover, multiple types of charge carriers are present and simple Hall resistivity measurements are only an estimate of the net carrier density. We show in Figure 2(c) the carrier density of BTS estimated for a single carrier model[13] from the Hall resistivity measurements at different positions in the boule at 16, 100 and 200 K. At high temperatures (200 and 100 K) the samples (A and B) from the first to freeze region dominantly possess *n*-type carriers, whereas rest of the crystal boule shows the dominance of *p*-type carriers. The carrier density at 100-200 K lies in the range of $10^{17}$-$10^{19}$ $cm^{-3}$. At 16 K the carrier densities is quite low, on the order of $\sim 10^{17}$ $cm^{-3}$ for nearly the entire boule. (At these low temperatures the surface states have an impact on the measured $R_H$.) For samples D, E and F, the middle of the boule, the carrier concentration is found to be of the order of ~ 8 x $10^{15}$ $cm^{-3}$; low for stoichiometry-compensated BTS. The last to freeze part of the crystal boule (samples F, G, and H) remains *p*-type at 16 K. The low carrier densities are a consequence of compensation of the electrons normally present by the holes created by the $Bi_{Te}$ antisite defects, which act as acceptors,[9] and indicates that the VBT-grown BTS crystal is of high quality.

Similarly, Figure 2(d) shows the carrier density of BTS:Sn evaluated at 16, 150 and 200 K. The behavior is quite different from that of the BTS boule. Even at 200 and 150 K, the (n-type) carrier density for the first 9 cm of the boule is of the order of $10^{17}$ $cm^{-3}$ or less. Sample C1 exhibits a carrier density at 150 K of the order of ~ 8 x$10^{14}$ $cm^{-3}$, which is very low for a small band gap semiconductor and for a material in the Tetradymite family. Thus, experimentally, Sn at the Bi site is found to act as an acceptor, resulting in electron compensation and ultimately *p*-type doping in BTS. The consistently low bulk carrier density over a large region of the boule depicts the high quality of the grown crystal and the suitability of Sn-doped BTS for observing the behavior of carriers in the topological surface states.

## B. Shubnikov de Haas (SdH) oscillations

Along with the high values of the bulk resistivity, both Bi-excess BTS and Sn-doped BTS display Shubnikov de Haas (SdH) oscillations from the surface states in measurements of the magnetoresistance. The surface nature of these oscillations was previously verified,[21,23] and their appearance attests to the high mobility of the topological surface state electrons in both types of crystals. Figure 3(a) and (b) display curves of the magnetoresistance for representative samples of both BTS and BTS:Sn crystals after a background subtraction. The quantum oscillations are observed only in samples cut from parts of the boules that are highly insulating (For example, section D1 in BTS:Sn and section D in BTS). The Sn-doped crystals show larger and more well-defined oscillation patterns than those usually observed in pure BTS, probably due to the reduced interference with the less conducting bulk. In both Figure 3(a) and (b), the insets display the Landau index plots, from which we derive the surface Fermi wave-vector $k_F$ and surface Fermi energy $E_F$. The Sn-doped BTS samples tested showed similar values for $k_F$ and $E_F$ ($k_F$ = 0.065 Å$^{-1}$, $E_F$ = 255 meV). Although the surface Fermi energy in Sn-doped BTS is higher than that typically observed in pure BTS crystals (180-190 meV), which can be seen from the higher frequency of the magnetoresistance oscillations in 1/B in Figure 3, (a measure of the size of the surface Fermi surface, which increases as one departs in energy from the surface Dirac point) it is inside the bulk energy gap.

## C. Angle resolved photoelectron spectroscopy

Figure 4 shows the ARPES data from pristine and slightly Cr-covered BTS:Sn surfaces. The topological surface state (TSS) on the pristine surface forms a small, circular electron Fermi pocket, similar in size to the one typically seen in $Bi_2Se_3$, with $k_F$ = 0.06 Å$^{-1}$. This is in agreement with what is observed for the sdH oscillations (observed on a different crystal, attesting to the crystal uniformity near the middle of the boule) corresponding to a surface electron density of $n \approx 2.9\text{x}10^{12}$ cm$^{-2}$. Its dispersion and the energy of the Dirac point ($E_D \approx$ 250 meV, again in agreement with the sdH measurements) below the chemical potential are also very similar to that seen for $Bi_2Se_3$. There are several significant differences. One is the position of the bulk valence- and conduction bands (BVB and BCB) with respect to the surface states: whereas in $Bi_2Se_3$ the Dirac point lays above the BVB maximum, in $Bi_{1.99}Sn_{0.01}Te_2Se$ it is slightly buried, as the BVB maximum is ≈ 50 meV higher. (This will be further discussed in reference to Figure 5). Also, in $Bi_2Se_3$ there are a significant number of electrons occupying the bottom of BCB near the center of the bulk Brillouin zone, $k_z$ = 0,[24] while for BTS:Sn the BCB remains unoccupied at any $k_z$. Only after additional electron doping through the adsorption of electron donors (i.e. Cr) at the surface, does the BCB become visible in the ARPES measurements. In Figure 4(d), we show the spectrum of the surface covered by roughly a tenth of a monolayer of Cr. In addition to an enlarged TSS, the bottom of BCB becomes partially occupied, with its minimum at ≈ 90 meV. Also, the BVB shifts down by ≈ 200 meV and a very flat band is seen, marked by an arrow. This feature was not seen in previous studies where various adsorbates were deposited on $Bi_2Se_3$.[25] The bulk band gap can be estimated to be approximately 300 meV from the ARPES data.

In Figure 5, which shows the detailed $k$-resolved electronic structure of pristine BTS:Sn, we further disentangle the bulk states from the topological surface states (TSS). Panel (a) shows the photoemission intensity at $E = E_D$ (the Dirac point energy) as a function of the in-plane momentum. Aside from the Dirac point at the zone center, petal-like protrusions are seen extending along the $\Gamma$ – M lines of the surface Brillouin zone (SBZ). Their maxima are ~ 60 meV higher in energy than $E_D$, as indicated in panel (b). Our photon-energy dependent studies indicate that these protrusions are bulk states. Panels (c) and (d) show the $k_z$-dependence of the ARPES intensity at $E = 0$ and at $E = E_D$, respectively, along the in-plane momentum line marked by the dashed line in panel (a). Panel (e) shows the $k_z$ dependence of intensity at the center of the SBZ. The intensity maps recorded at photon energies from 50 to 76 eV (in steps of 2 eV) are converted into $k$-space by using the inner potential V = 10 eV in the approximation for $k_z = \frac{1}{h}\sqrt{2m(E_k cos^2(\theta) + V)}$, where $E_k$ is the kinetic energy of a photo-electron. The straight features at $k_x = \pm 0.06$ Å$^{-1}$ in Figure 5(c) represent the Fermi surface of the TSS. The bulk character of the petal-like streaks from panel (a) is confirmed here in panels (d) and (e) as they clearly disperse with $k_z$, on both sides of the dispersionless Dirac point. They show the periodicity of ≈ 0.62 Å$^{-1}$, suggesting that the relevant period in real space in the $z$ direction is ≈ 1 nm, close to the thickness of one quintuple layer. It is important to note that the BCB does not appear in either panel (c) or (e) at any point in the bulk BZ, implying that the TSS are the only highly occupied states at the Fermi level (therefore having the potential to dominate the low-temperature charge transport in BTS:Sn). However, due to the energy of the Dirac point relative to the top of the BVB, even if the chemical potential can be brought to the Dirac point in BTS:Sn by an applied field or chemical doping with the intention of entering the "Dirac" regime, there would be substantial interference by the bulk holes from the BVB.

Figure 6 shows the bulk-hole concentration deduced from the six petal-like features and the surface-electron concentration from the TSS as functions of chemical potential (relative to $E_D$, which is 60 meV below the BVB maximum). Due to uncertainty in the measured $k_z$ dependence, the volume of the bulk Fermi pockets can be only approximately calculated. Therefore, we show the two extreme cases: the blue line is an approximation in which there is no warping of the bulk FS with $k_z$ (i.e. for a 2D bulk electronic system) and the red line assumes completely closed 3D bulk Fermi surfaces (i.e. for a 3D bulk electronic system). The actual carrier concentration lies between these two limits. The rapid increase in *p*-type bulk carrier concentration as a function of energy on entering the BVB indicates that carrier concentrations in the $10^{14}$ $cm^{-3}$ regime, as obtained for our vertical Bridgeman grown Sn-doped BTS crystals, are exceptional. This analysis shows that at only 2 meV inside the BVB, the bulk hole concentration will already be in the $10^{17}$ range, implying either that $E_F$ in our VBT-grown Sn-doped BTS crystals is exquisitely well placed over a crystal length of many centimeters due to control of the defect chemistry, or that some special mechanism is present to pin the Fermi energy. Since the dopant concentration cannot possibly be so well controlled over such long distances in the crystal, the latter must be the case. We attribute this behavior to the fact that Sn is a special impurity in BTS (see below).

### D. Electronic structure and density of states

To help explain the significantly higher resistivity and lower carrier concentrations in Sn doped BTS crystals when compared to those grown with excess Bi, electronic structure calculations were performed on three systems: BTS with a 4% Bi on Te antisite defect, BTS with a 4% Sn on Bi site impurity, and stoichiometric BTS. The electronic structures of these three systems are shown in Figure 7. It can be clearly seen that both impurities create a defect state that lies in the vicinity of the top of the valence band, confirming their nature as *p*-type dopants. Both impurity levels are neither completely full nor completely empty, indicating the impurities' ability to absorb extra electrons or holes without shifting the Fermi energy very much. The Sn impurity level, however, has both a much smaller bandwidth and is significantly less hybridized with the rest of the valence bands than the Bi antisite impurity level. It is also very flat in E vs. *k*, reflecting its highly localized character. A comparison between the DOS of undoped BTS and the Sn and Bi antisite doped systems (Figure 7) shows this in further detail. The Sn-doped case has a small peak in the DOS centered right near 0 eV, or at the very top of the valence band. The Bi antisite case has a small peak at this point too, but has a more prominent extra contribution to the DOS around -0.2 eV from $E_F$. The Sn doped case therefore clearly has a more well-separated impurity level. It is notable that the calculation was done with 4% Sn, compared to the 0.5% Sn in the real crystals. One thus expects that the real case would have a sharper, even more well-separated and localized state than the calculated case. This result is consistent with previous studies on Sn doped $Bi_2Te_3$,[26,27] where Sn is known to create a resonant level.[19,20] In $Bi_2Te_3$, the Sn impurity creates a resonant state about 15 meV below the top of the valence band. The difference in Se and Te electronegativities indicates that is reasonable that on going from $Bi_2Te_3$ to $Bi_2Te_2Se$ the relative energy levels change such that the Sn impurity band now appears in the gap near the top of the valence band.

The Sn impurity level around $E_F$ is mainly composed of Sn 5*s* orbitals. Given that this state is partially filled in BTS:Sn, this implies that the Sn impurities have an oxidation state somewhere between 4+ and 2+ in this compound. Given that Sn in a Te octahedron is often 2+, as in SnTe, and in a Se octahedron is often 4+, as in $SnSe_2$, this makes sense from a chemical perspective. A likely scenario is that the Sn impurities can be either 4+ or 2+, and that extra electrons or holes can be absorbed to create either more Sn 4+ or Sn 2+, i.e. that the Sn forms a resonant state. The electronic structure calculation gives an "average" of this scenario. Given the flatness of the band, the Sn 5s electrons are localized and not conductive. Thus the Sn dopants act as a carrier concentration "buffer", absorbing electrons or holes into a localized band; the other defects present that would normally add conducting electrons or holes to the system instead simply either reduce or oxidize the Sn impurities from 4+ to 2+ and vice versa. In this way, Sn impurities are a special defect, whose use helps to position Sn-doped BTS as an excellent candidate for studying surface state transport in topological insulators.

## IV. CONCLUSION

We have described a two-zone furnace that is suitable for growing high quality bulk crystals of BTS. High carrier compensation in the crystals was achieved by either excess of Bi or Sn doping, and crystals were grown with the dominance of *p*-type carriers, which is different from the case of nominally stoichiometry BTS, which is *n*-type. The grown crystals are found to be highly resistive over a large range of temperature and the resistivity reaches values as high as 10 **Ω**-cm at 150 K and below for the Sn-doped crystals. The thermal activation energy for a large section of the Sn-doped crystal was found to be on the order of ~140 meV, indicating that the Fermi level is in the bulk band gap. The lowest bulk carrier densities observed for the grown crystals were on the order of $10^{15}$ $cm^{-3}$ and $10^{14}$ $cm^{-3}$ respectively for the Bi excess and Sn doped crystals. Electronic structure calculations show that the very low carrier concentration and high resistivity are due to the special nature of the Sn impurity in BTS, specifically the resonant level behavior of Sn 5*s* orbitals at the top of the bulk valence band. Also, the existence of high quality SdH oscillations from the topological surface states in our Sn-doped BTS implies that the crystals grown through the vertical Bridgeman method are ideal for transport study of Dirac surface states. Employment of the crystal growth method described here will facilitate future studies of the transport properties of the surface electronic states of topological insulators.

## ACKNOWLEDGEMENTS

This research was supported by the ARO MURI on Topological Insulators, grant W911NF-12-1-0461, ARO grant W911NF-11-1-0379, and by the US Department of Energy (DOE) under Contract No. DE-AC02-98CH10886. ALS is operated by the US DOE under Contract No. DE-AC03-76SF00098.

## FIGURE CAPTIONS

**FIG. 1.** (Color on line) (a) Main panel ρ vs. T for samples from different positions of a Bi-excess $Bi_2Te_2Se$ crystal boule, designated by capital letters as described in text. The inset shows log ρ vs. 1/T in the high temperature range, employed to calculate the transport activation energy. (b) Main panel ρ vs. T for samples from different positions in the Sn-doped $Bi_2Te_2Se$ crystal boule, designated by capital letters as described in text. The inset shows log ρ vs. 1/T plots in the high temperature range, employed to calculate the transport activation energy.

**FIG. 2.** (Color on line) (a) and (b) are the $R_H$ vs. T plots for different parts of the BTS and BTS:Sn crystal boules. Log of carrier concentration evaluated at 16, 100 and 200 K plotted along the length of the (c) BTS and (d) BTS:Sn crystal boules. The lines are a guide to the eye. FIRST and LAST represent the first and last to freeze ends of boule, respectively.

**FIG. 3.** (Color on line) The SdH oscillation data from magnetoresistance (MR) measurements of (a) BTS and (b) BTS:Sn. The MR data were measured at 4.5K and smooth background curves have been subtracted to extract the SdH oscillation data. The insets are the Landau index plots for both cases. From the slopes, we find that in BTS $k_F$=0.048$Å^{-1}$ and $E_F$=190meV, while $k_F$ is 0.065$Å^{-1}$ and $E_F$ is 255meV for BTS:Sn.

**FIG. 4.** Photoemission Spectra from BTS:Sn. (a) The Fermi surface (b) ARPES spectrum along the Γ–*K* line in the SBZ. (c) ARPES spectrum along the Γ – M line. (d) ARPES spectrum measured along the Γ–M line after depositing ~0.1 monolayers of Cr. The arrows indicate the positions of BVB maximum and BCB minimum. Dotted and solid lines indicate shifts of bulk bands and the surface Dirac point, respectively, relative to the pristine surface. All the spectra were recorded at 70 eV photon energy.

**FIG. 5.** (Color on line) Surface and bulk electronic structure of BTS:Sn. (a) ARPES intensity at the binding energy corresponding to the Dirac point ($E_D$ = -250 meV) as a function of the in-plane momentum. The dashed oval-like contour outlines the cross-section of one of the six top segments of the bulk valence band (BVB). (b) ARPES spectrum along the solid blue line in (a). The arrow and the green line indicate the BVB maximum and the energy of the Dirac Point, respectively. (c) The Fermi surface (E = 0) intensity map as a function of the in-plane momentum along the dashed line in (a) and $k_z$. (d) Same as (c), but at $E = E_D$. (e) Intensity at the Γ point ($k_x = k_y = 0$) of the SBZ as a function of $k_z$. The data in (a) and (b) were recorded with 70 eV photons, while (c-e) were measured at photon energies ranging from 50 to 76 eV in 2 eV steps.

**FIG. 6.** (Color on line) The hole concentration deduced from the six petal-like features in the bulk valence band (figure 7a) as a function of chemical potential for energies below the Dirac point energy ($E_D$) for BTS:Sn. The blue line is an approximation in which there is no warping of the bulk FS with $k_z$ (i.e. for a 2D bulk electronic system) and the red line assumes completely closed 3D bulk Fermi surfaces with the approximation $k_{zF} = \sqrt{k_{xF}k_{yF}}$ (i.e. for a 3D bulk electronic system). The surface electron concentration for the topological surface state is also shown (dashed line). Note that the spin degeneracy is 1 (2) for the surface (bulk) states.

**FIG. 7.** (Color on line) (a) The calculated wavevector-dependent electronic band structure in the vicinity of $E_F$ for Bi excess $Bi_2Te_2Se$. The states dominated by contributions from the $Bi_{Te}$ antisite defect, near the top of the valence band that is due to the normal bulk states, are shown by the use of circles whose size is proportional to the defect contribution; (b) The calculated wavevector-dependent electronic band structure in the vicinity of $E_F$ for Sn-doped $Bi_2Te_2Se$. The states dominated by contributions from the Sn defect, near to but distinct from the top of the valence band that is due to the normal bulk states, are shown by the use of circles whose size is proportional to the defect contribution; (c) the corresponding electronic structure for stoichiometric $Bi_2Te_2Se$ (d) comparison of the calculated electronic densities of states for stoichiometric $Bi_2Te_2Se$, Bi-excess $Bi_2Te_2Se$, and Sn-doped $Bi_2Te_2Se$.

Figure 1

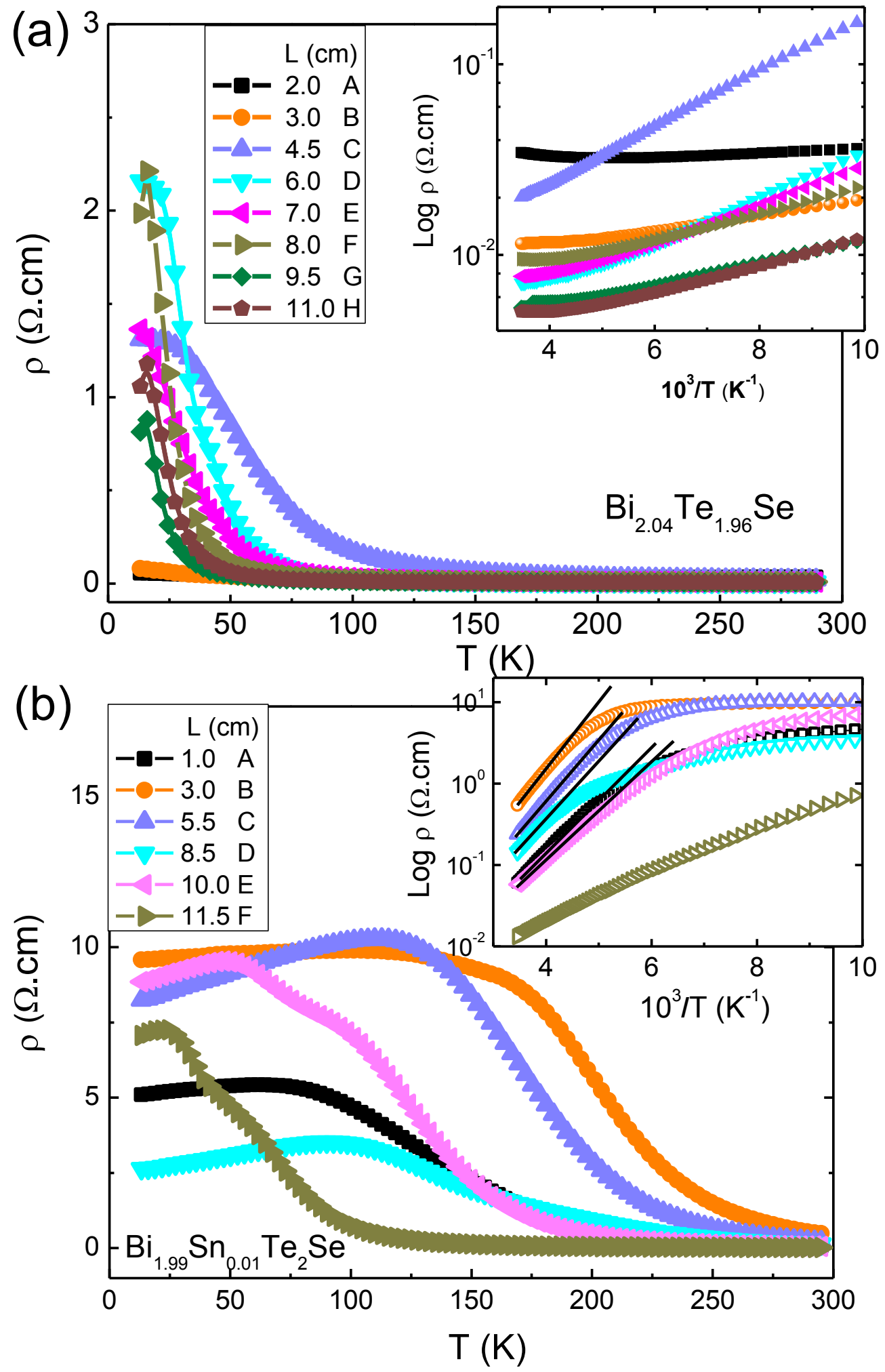

Figure 2

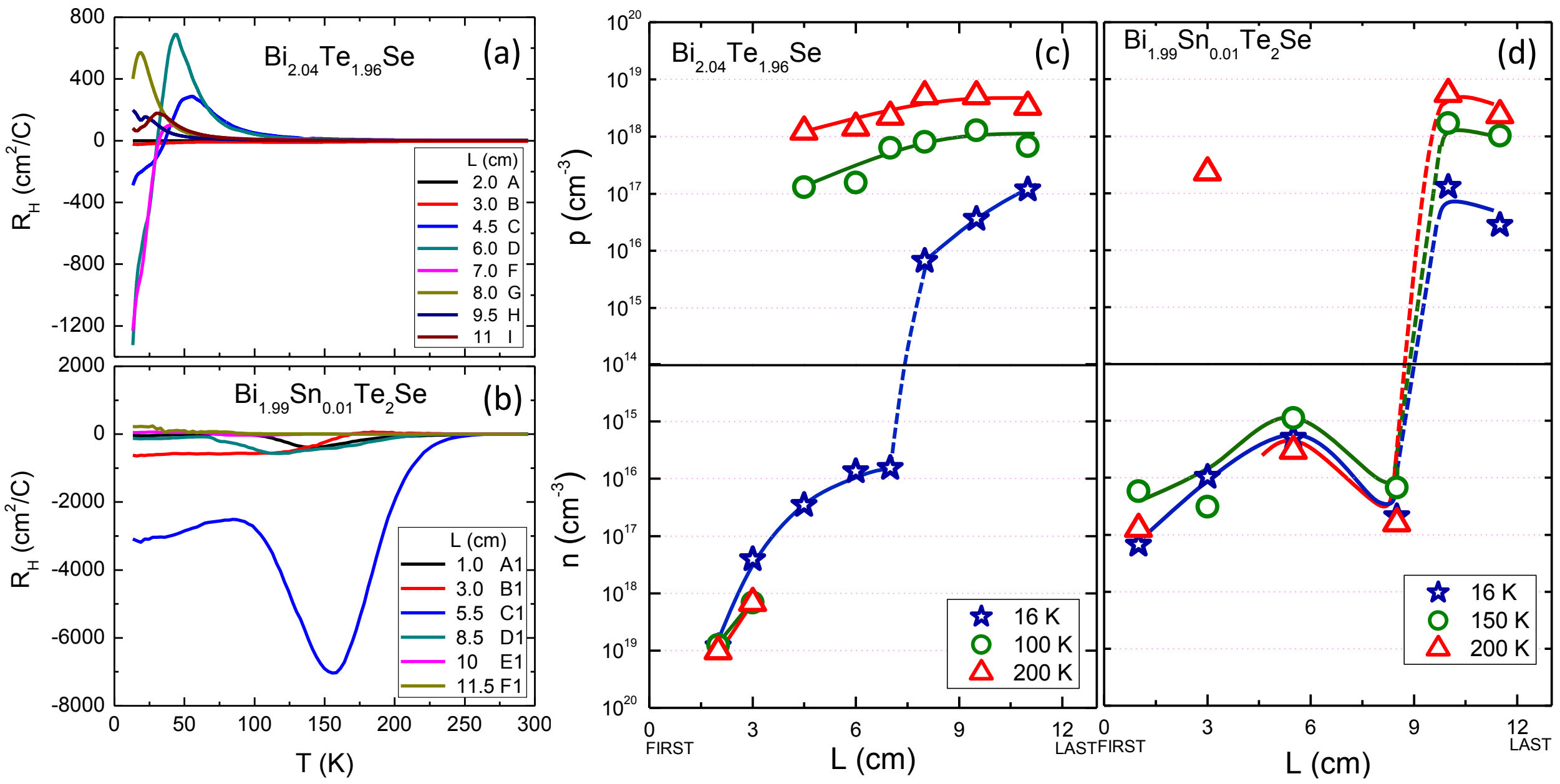

Figure 3

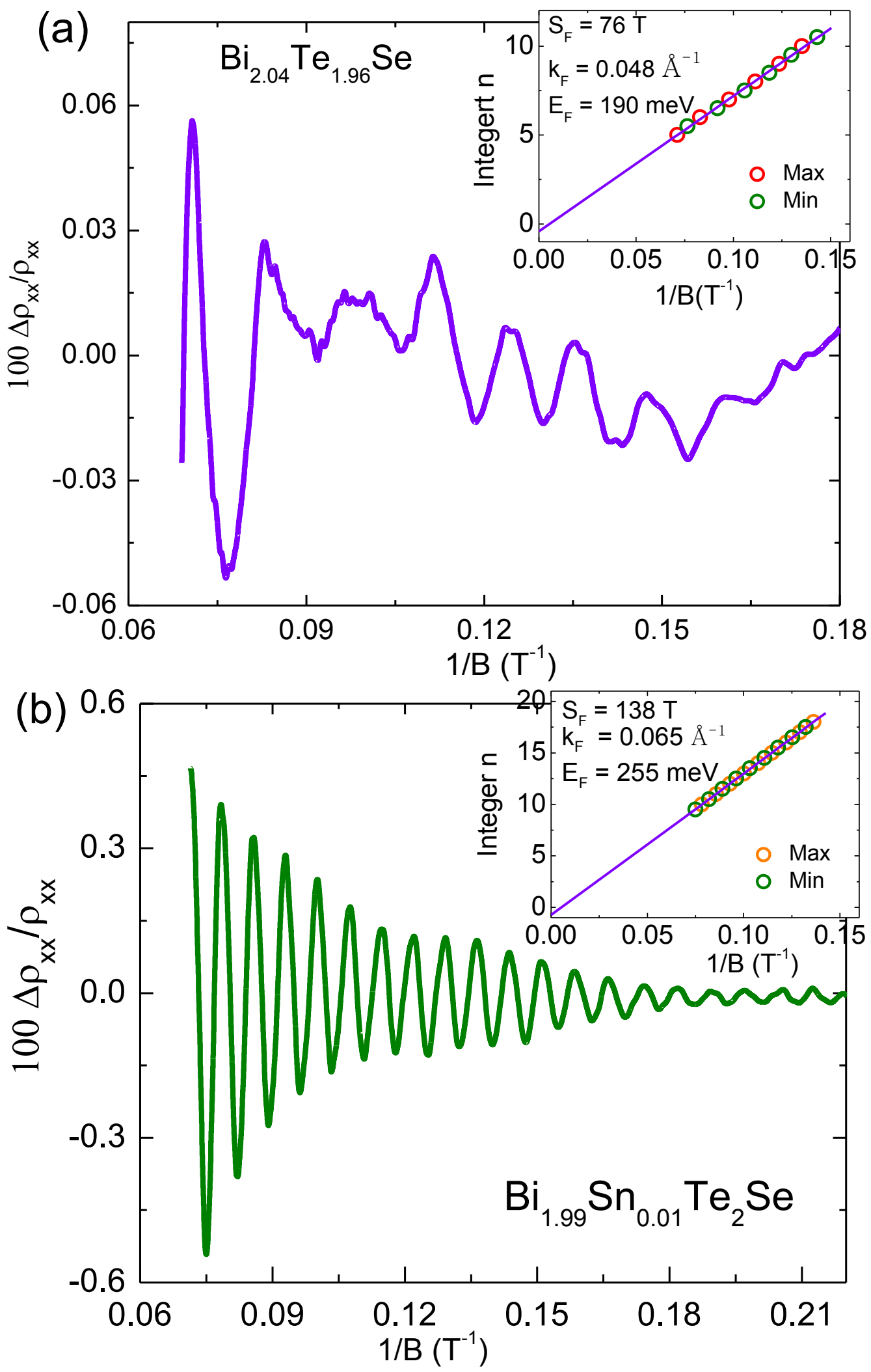

(a)
Bi2.04Te1.96Se
100 Δρxx/ρxx
1/B (T-1)
SF = 76 T
kF = 0.048 Å-1
EF = 190 meV
Integert n
1/B(T-1)
Max
Min
(b)
100 Δρxx/ρxx
1/B (T-1)
Bi1.99Sn0.01Te2Se
SF = 138 T
kF = 0.065 Å-1
EF = 255 meV
Integer n
1/B (T-1)
Max
Min

Figure 4

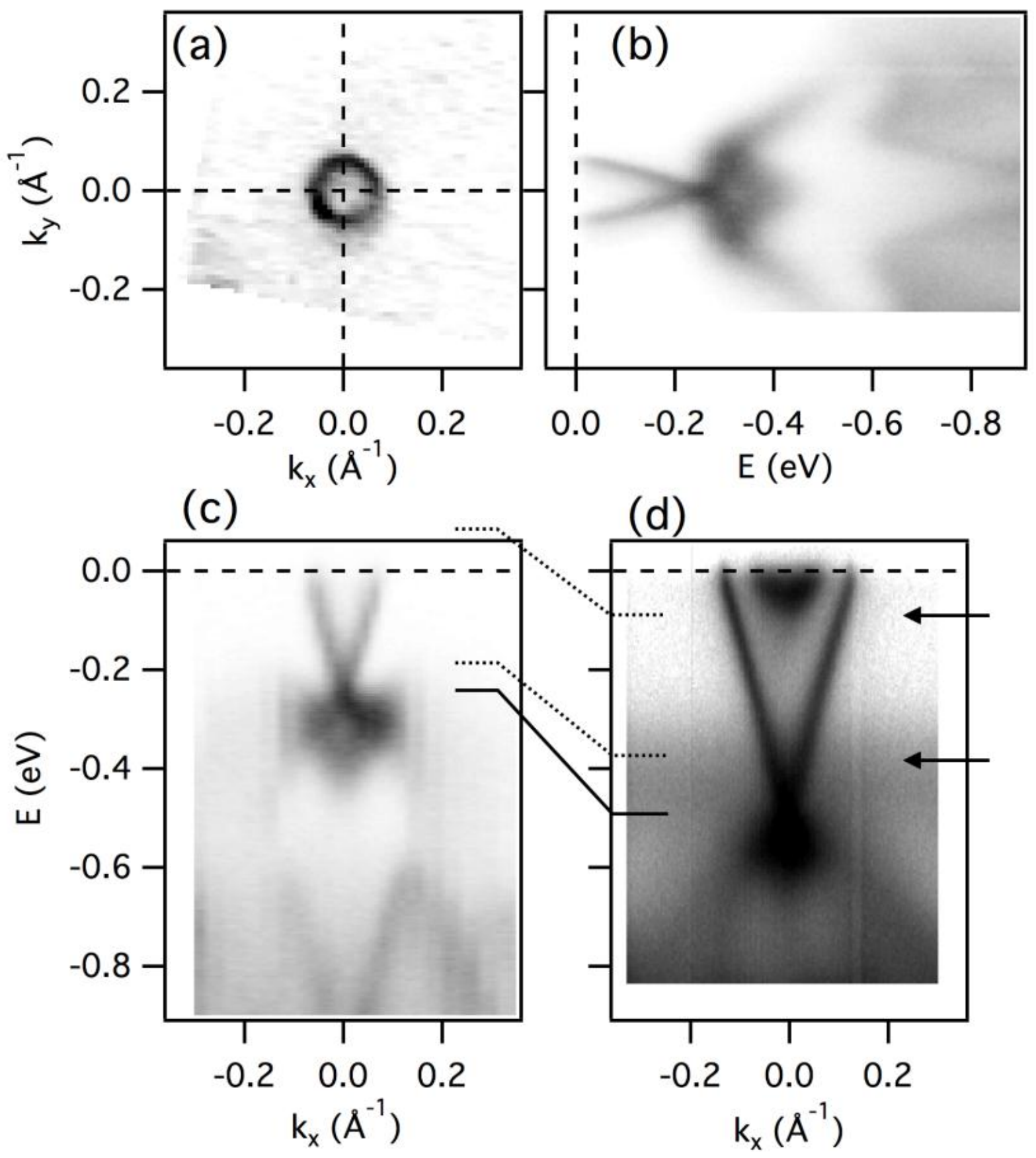

Figure 5

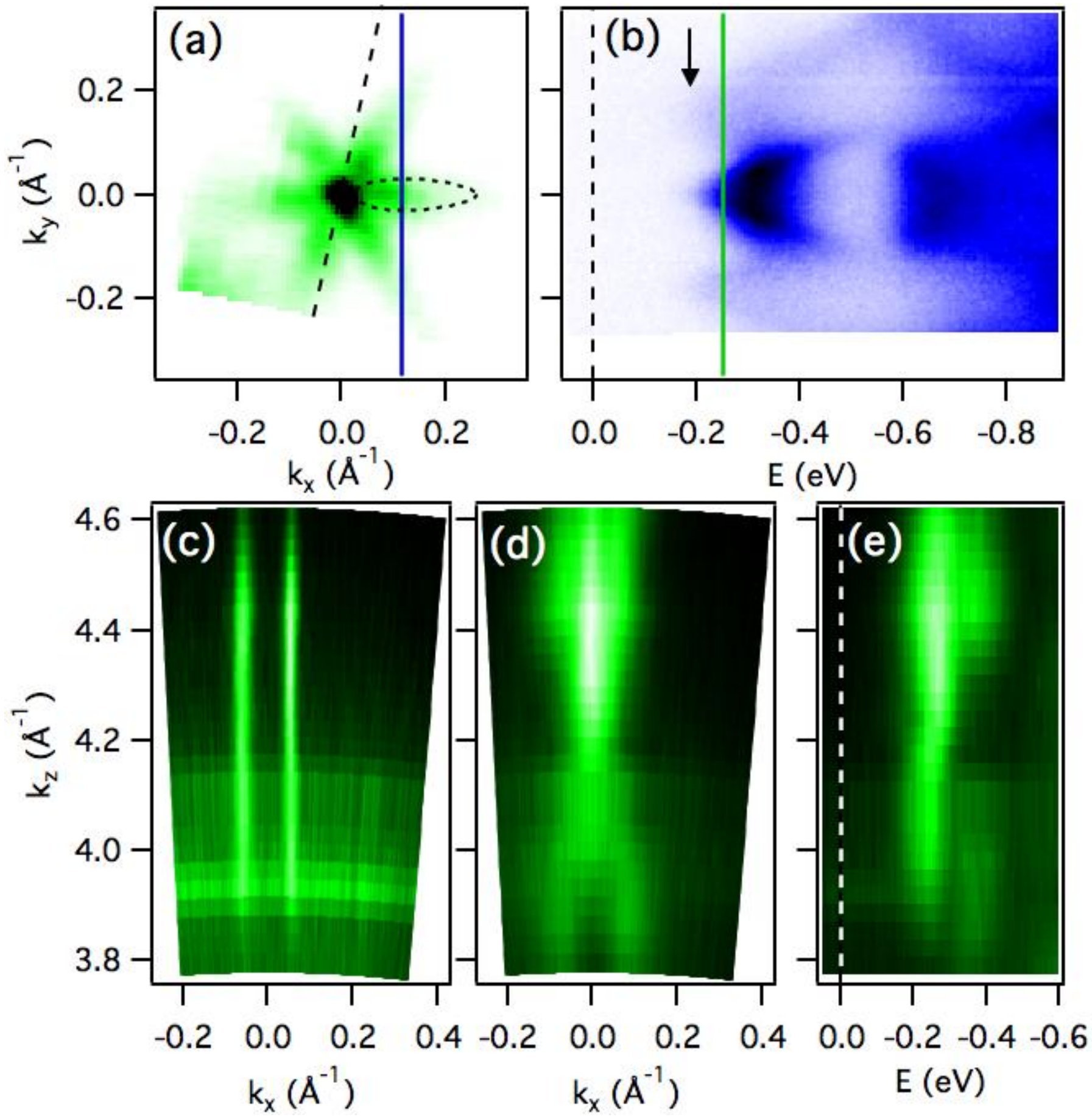

Figure 6

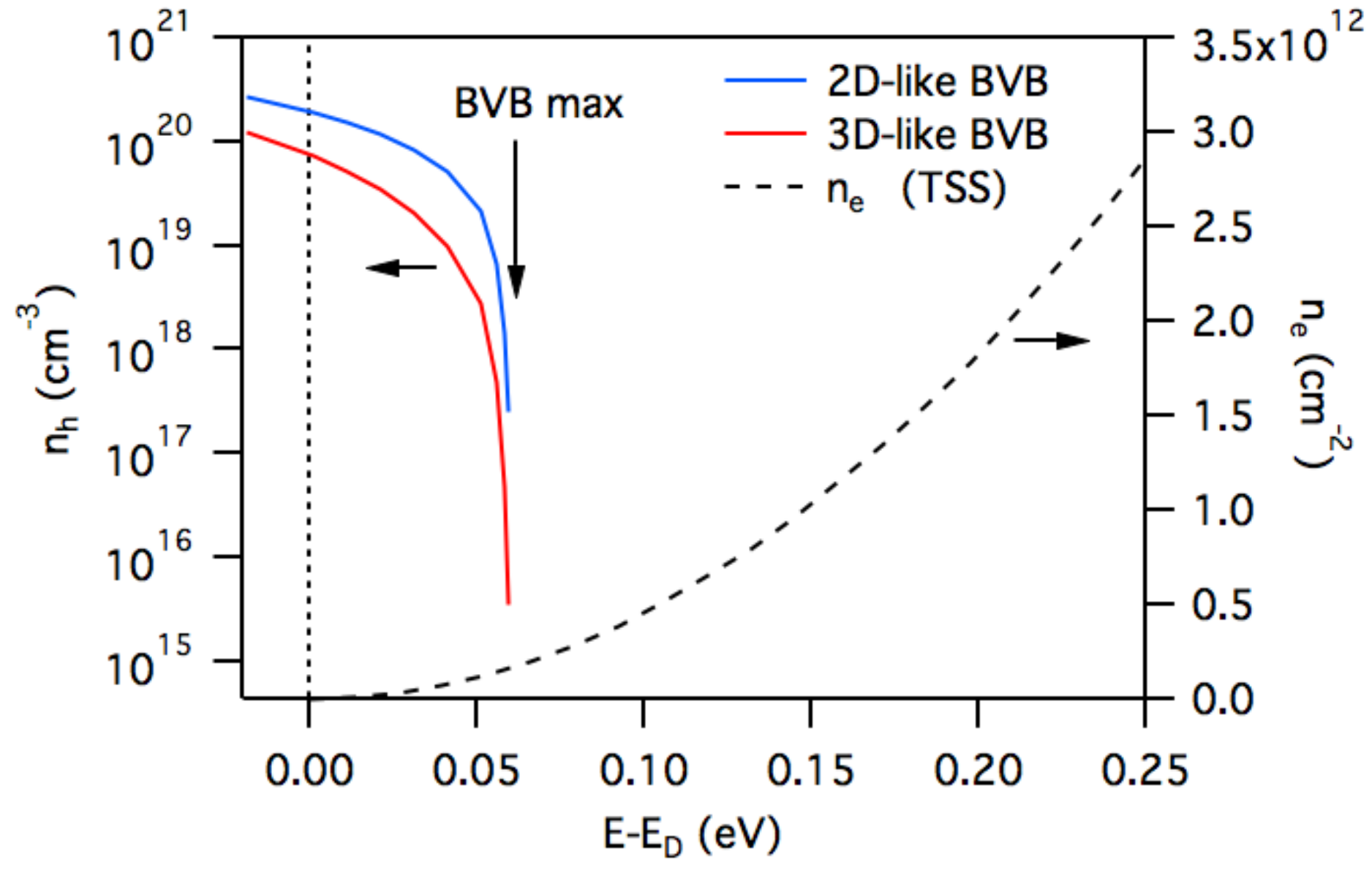

Figure 7

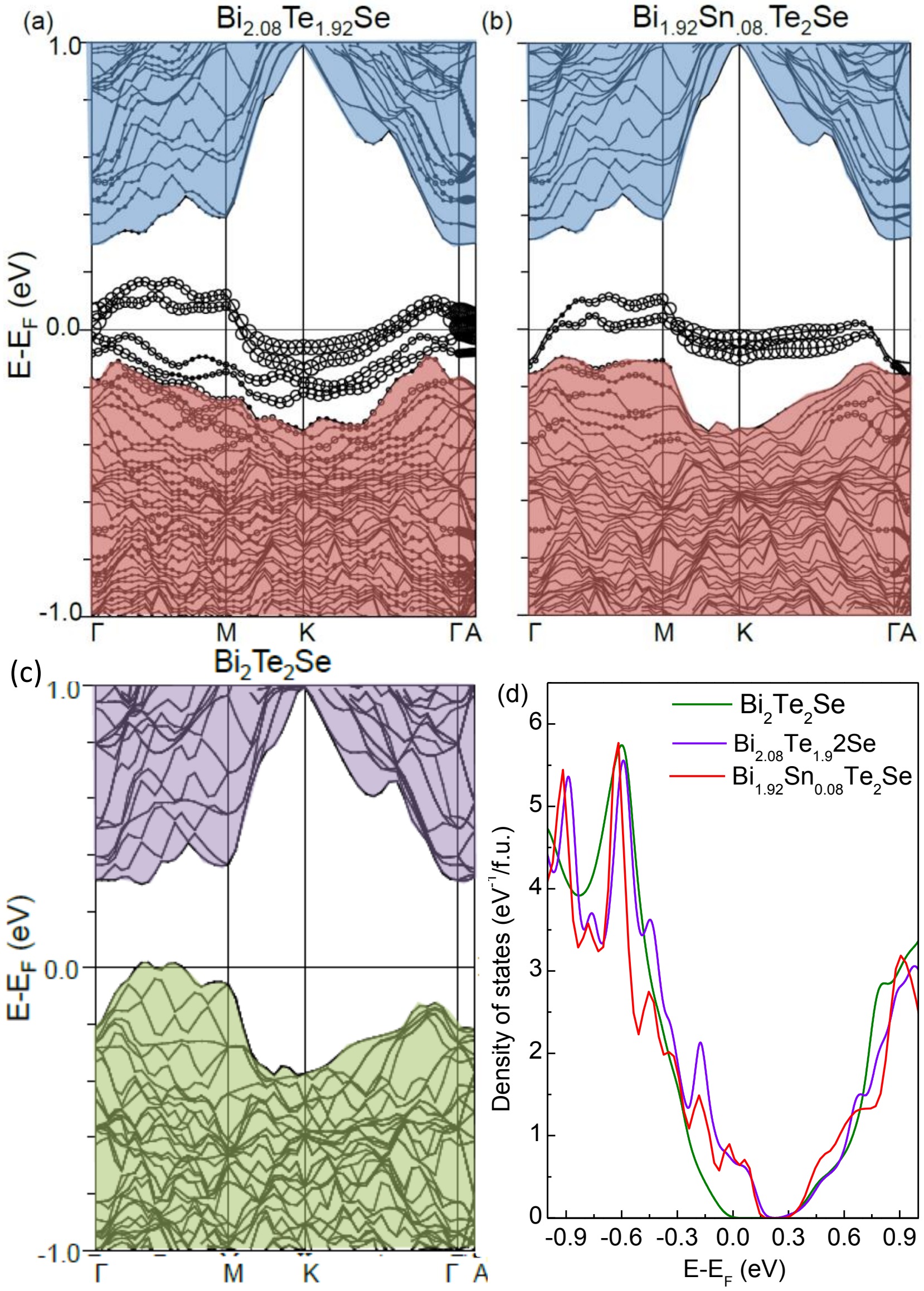